\documentclass[aps,showpacs,superscriptadress,preprint]{revtex4}
\usepackage{graphicx}

\begin{document}
\title{Liquid-gas Phase Transition in Strange Hadronic Matter with Weak Y-Y Interaction}

\author{Li Yang$^1$\footnote{georgeyangli@yahoo.com.cn}, Shao Yu Yin$^1$
,Wei Liang Qian$^1$\footnote{wlqian@fudan.edu.cn} and Ru-Keng
Su$^{2,1}$\footnote{rksu@fudan.ac.cn}} \affiliation{
\small 1. Department of Physics, Fudan University,Shanghai 200433, China\\
\small 2. CCAST(World Laboratory), P.O.Box 8730, Beijing 100080, China\\
}

\begin{abstract}
The liquid-gas phase transition in strange hadronic matter is
reexamined by using the new parameters about the $\Lambda -
\Lambda$ interaction deduced from recent observation of
$^{6}_{\Lambda\Lambda}He$ double hypernucleus. The extended
Furnstahl-Serot-Tang model with nucleons and hyperons is utilized.
The binodal surface, the limit pressure, the entropy, the specific
heat capacity and the Caloric curves are addressed. We find that
the liquid-gas phase transition can occur more easily in strange
hadronic matter with weak Y-Y interaction than that of the strong
Y-Y interaction.
\end{abstract}

\maketitle

\section{Introduction}

Strangeness opens a new dimension to nuclear physics.  Nuclear
system with strangeness has many implications in astrophysics and
cosmology and thus arouses much research interest.  Nuclear
systems with strangeness can be categorized into two species:
strange quark matter(or strangelet) and strange hadronic
matter(SHM). Quantum chromodynamics predicts that the
strangenss-rich system could be strange quark matter or
strangelet, consisting of \emph{u}, \emph{d} and \emph{s} quarks.
Strangelet has been a hot research topic for both theoretical
physics and experimental physics, but there is still no convincing
experimental result to confirm its existence. The other possible
existence of strangeness is SHM, in which quarks are localized
within nucleons and hyperons.  The research of SHM is attractive
because of the complexity of its structure and ingredients,
especially, the phase transition including the quark deconfinement
phase transition and liquid-gas(L-G) phase transition.

This paper evolves from an attempt to study the L-G phase
transition in SHM at finite temperature. In a previous
paper\cite{Yang:2003up}, we have studied the possibility of an L-G
phase transition in SHM utilizing the extended
Furnstahl-Serot-Tang(FST) model. After investigating the
mechanical and chemical stabilities of SHM, we found a critical
pressure on the binodal surface for $T=13MeV$ and a limit pressure
on the binodal surface for lower temperature.  After examining the
continuities of entropy and specific heat capacity, we found the
L-G phase transition in SHM is of second order.  Employing the
chiral $SU(3)$ quark mean field model, the same conclusion has
also been drawn in ref.\cite{Wang:2004gy}.

Recently, Takahashi \emph{et al.}\cite{Takahashi:2001nm} observed
the $\Lambda - \Lambda$ energy in a $^{6}_{\Lambda\Lambda}He$
double hypernucleus. Their observation deduced the $\Lambda -
\Lambda$ energy to be $\Delta
B_{\Lambda\Lambda}=1.01\pm0.20^{+0.18}_{-0.11}MeV$, which is much
smaller than the previous estimation $\Delta
B_{\Lambda\Lambda}\cong4\sim5MeV$ from earlier
experiments\cite{Danysz:1963,Dalitz:1989,Prowse:1966,Aoki:1991ip,Dover:1991kf}.
This leads to a significant decrease of potential
$V_{\Lambda}^{(\Lambda)}$ from about $20MeV$ to
$5MeV$\cite{Song:2003xe,Qian:2004td}. The new data imply that the
hyperon-hyperon interaction(Y-Y interaction) can be weaker than
known before\cite{Song:2003xe}.  We employed two of many
candidates of phenomenological SHM
models\cite{Zhang:2000ug,Qian:2003ct,Qian:2004td,Ikeda:1985um,Schaffner:1993qj,
Schaffner-Bielich:2000wj,Schulze:1998jf,Zhang:2003fn}, namely,
extended FST model\cite{Zhang:2000ug,Qian:2003ct,Qian:2004td} and
modified quark meson coupling(MQMC) model\cite{Song:2003xe}, and
the weak Y-Y interaction to reexamine the stability of SHM.  We
came to a conclusion that, while the system with the strong Y-Y
interaction and in a quite large strangeness fraction region is
more deeply bound than the ordinary nuclear matter due to the
opening of new degrees of freedom, the system with weak Y-Y
interaction is rather loosely bound compared to the
latter\cite{Song:2003xe}.  This conclusion is not model-dependent
on the extended FST model or MQMC model and true for both zero and
finite temperature\cite{Qian:2004td}.

Noting that the stability affects on the L-G phase transition
directly, the stability of SHM with weak Y-Y interaction is quite
different from that with strong Y-Y interaction.  It is of
interest to investigate the influence of this difference on the
L-G phase transition in SHM.  To employ the extended FST model and
reexamine the L-G phase transition in SHM with weak Y-Y
interaction is the main purpose of this paper.  We will show that
there is still a second order L-G phase transition in SHM with
weak Y-Y interaction, but the binodal surface, limit pressure and
the equation of state are very different from those of strong Y-Y
interaction.

The paper is organized as follows.  In Sec. II, the extended FST
model is briefly introduced and the parameters for strong and weak
Y-Y interactions are given.
 In Sec. III we reexamine the L-G phase transition in SHM with weak Y-Y interaction.
The last section is the summary of the main results.

\section{The extended FST model}

In refs.\cite{Zhang:2000ug,Qian:2003ct,Qian:2004td}, we extended
the original FST model to include not only nucleons and $\sigma$,
$\omega$ mesons, but also $\Lambda$, $\Xi$ hyperons and strange
mesons $\sigma^*$ and $\phi$.  In ref.\cite{Yang:2003up}, we used
this model to study the L-G phase transition in SHM.  The details
of this model can be found in refs.\cite{Yang:2003up,Zhang:2000ug,
Qian:2003ct,Qian:2004td}. Hereafter, we only give brief formulae,
which are necessary for discussing the L-G phase transition.

In mean-field approximation, the Lagrangian of the extended FST
model is presented as follows:

\begin{eqnarray}
{\cal L}_{MFT} &=&\bar{\psi}_{N}(i\gamma ^{\mu }\partial _{\mu }
-g_{\omega N}\gamma ^{0}V_{0}-M_{N}+g_{sN}\sigma _{0})\psi _{N}
\nonumber \\ &&+\bar{\psi}_{\Lambda }(i\gamma ^{\mu }{\partial }
_{\mu }-g_{\omega \Lambda }\gamma ^{0}V_{0}-g_{\phi \Lambda }
\gamma ^{0}\phi _{0}-M_{\Lambda }+g_{s\Lambda }\sigma _{0}
+g_{\sigma ^{\ast }\Lambda }\sigma _{0}^{\ast })\psi _{\Lambda }
\nonumber \\ &&+\bar{\psi}_{\Xi }(i\gamma ^{\mu } {\partial }_{\mu
}-g_{\omega \Xi }\gamma ^{0}V_{0}-g_{\phi \Xi } \gamma ^{0}\phi
_{0}-M_{\Xi }+g_{s\Xi }\sigma _{0}+g_{\sigma ^{\ast } \Xi }\sigma
_{0}^{\ast })\psi _{\Xi }  \nonumber \\ &&+\frac{1}{2} \left(
1+\eta \frac{\sigma _{0}}{S_{0}}\right) m_{\omega }^{2}V_{0}^{2}
+\frac{1}{4!}\zeta \left( g_{\omega N}V_{0}\right) ^{4}+\frac{ 1
}{2} m_{\phi }^{2}\phi _{0}^{2}-\frac{1}{2}m_{\sigma ^{\ast
}}^{2}\sigma ^{\ast ^{2}}
  \nonumber \\ &&-H_{q}\left( 1-\frac{\sigma _{0}}{S_{0}}\right) ^{4/d}
  \left[ \frac{1}{d} ln\left( 1-\frac{\sigma _{0}}{S_{0}}\right) -\frac{1}{4}\right]
\end{eqnarray}
$ g_{ij}$ are the coupling constants of baryon $j$ to meson $i$
field.

By using the standard technique of statistical mechanics, the
thermodynamic potential $\Omega $ is obtained
\begin{eqnarray}
\Omega &=&V\{H_{g}[(1-\frac{\sigma
_{0}}{S_{0}})^{\frac{4}{d}}(\frac{1}{d} \ln (1-\frac{\sigma
_{0}}{S_{0}})-\frac{1}{4})+\frac{1}{4}]  \nonumber \\
&&-\frac{1}{2}(1+\eta \frac{\sigma _{0}}{S_{0}})m_{\omega
}^{2}V_{0}^{2}-  \frac{1}{4!}\zeta (g_{\omega
N}V_{0})^{4}-\frac{1}{2}m_{\phi }^{2}\phi
_{0}^{2}+\frac{1}{2}m_{\sigma ^{\ast }}^{2}\sigma _{0}^{\ast
^{2}}\}  \nonumber \\ &&-2k_{B}T\{\sum_{i,{\bf k}}\ln
{[1+e^{-\beta (E_{i}^{\ast }(k)-\nu _{i}}] } +\sum_{{i,{\bf
k}}}\ln {[1+e^{-\beta (E_{i}^{\ast }(k)+\nu _{i})}]}\}
\end{eqnarray}
where $\beta \ $is the inverse temperature and $V$ is the volume
of the system

\begin{equation}
E_{i}^{\ast }(k)=\sqrt{M_{i}^{\ast 2}+k^{2}}
\end{equation}
The effective masses of the hyperons and nucleons are
\begin{equation}
M_{i}^{\ast }=M_{i}-g_{si}\sigma _{0}-g_{\sigma ^{\ast }i}\sigma
_{0}^{\ast }\qquad(i=\Lambda ,\Xi ),
\end{equation}
\begin{equation}
M_{i}^{\ast }=M_{i}-g_{si}\sigma _{0}\qquad(i=N)
\end{equation}
The mean-field values $\phi _{0}$, $V_{0}$ , $\sigma _{0}$ and
$\sigma _{0}^{\ast }$ are determined by the corresponding extreme
conditions of the thermodynamic potential.

The  baryon densities $\rho _{Bi}$ is given by
\begin{equation}
\rho _{Bi}=\left\langle {\psi}^{+}_{i}\psi _{i}\right\rangle
=\frac{g_{i}}{ \pi ^{2}}\int dkk^{2}\left[ n_{i}\left( k\right)
-\bar{n}_{i}\left( k\right) \right]
\end{equation}
where $g_{i}=2$ for $i=N$ or $\Xi $, $g_{i}=1$ for $i=\Lambda $.
The baryon and anti-baryon distributions are, respectively,
expressed as
\begin{equation}
n_{i}(k)=\{exp[\beta (E_{i}^{\ast }(k)-\nu _{i})]+1\}^{-1}
\end{equation}
and
\begin{equation}
\overline{n}_{i}(k)=\{exp[\beta (E_{i}^{\ast }(k)+\nu _{i})]+1\}^{-1}
\end{equation}
$\nu _{i}$ are related to chemical potential $\mu _{i}$ by
\begin{eqnarray}
\mu _{N} &=&\nu _{N}+g_{\omega N}V_{0},  \nonumber \\
\mu _{\Lambda } &=&\nu _{\Lambda }+g_{\omega \Lambda }V_{0}+g_{\phi \Lambda
}\phi _{0},  \nonumber \\
\mu _{\Xi } &=&\nu _{\Xi }+g_{\omega \Xi }V_{0}+g_{\phi \Xi }\phi
_{0}.
\end{eqnarray}

Since the system has equal number of protons and neutrons, and
equal number of $\Xi^0$ and $\Xi^-$, the chemical equilibrium
conditions for the reactions $\Lambda +\Lambda \rightleftharpoons
n +\Xi^0$ and $\Lambda +\Lambda \rightleftharpoons p +\Xi^-$ are
unified as
\begin{equation}
2 \mu_\Lambda=\mu_N +\mu_\Xi.
\end{equation}
Eq.(10) implies that only two components of $N$, $ \Lambda $ and $
\Xi $ are independent.  The SHM is regarded as a two-component
system, for instance, the nucleon component and the $\Xi$
component.  The strangeness fraction$f_S$ is introduced as
\begin{equation}
 f_{S}\equiv
\frac{\rho _{B\Lambda }+2\rho _{B\Xi }}{\rho _{B}}
\end{equation}
which plays the similar role as that of the asymmetric parameter $
\alpha = ( \rho_{n} - \rho_{p}) / (\rho_{n} + \rho_{p})$ in the
asymmetric nuclear matter.  We can use the same method as that of
refs.\cite{Yang:2003up,Muller:1995ji,Qian:2000fq,Qian:2001dw,Qian:2002rp,Qian:2002kj}
to address the L-G phase transition.

Following the usual procedure of statistical physics, we can
calculate the other thermodynamic quantities from thermodynamic
potential $\Omega $.  For example, the pressure and entropy
density are calculated by formulas $p=-\Omega /V$ and $
S/V=-(\partial \Omega /\partial (1/\beta ))_{V,\mu
_{i}}/V=(\partial p/\partial (1/\beta ))_{V,\mu _{i},\sigma
_{0},V_{0},\phi _{0},\sigma _{0}^{\ast }}$.  The results are
expressed as below
\begin{eqnarray}
p &=&\sum_{i}\frac{g_{i}}{6\pi ^{2}}\int dk\frac{k^{4}}{E_{i}^{\ast }(k)}
[n_{i}(k)+\overline{n}_{i}(k)]-H_{q}\left\{ \left( 1-\frac{\sigma _{0}}{S_{0}
}\right) ^{\frac{4}{d}}\left[ \frac{1}{d}ln\left( 1-\frac{\sigma _{0}}{S_{0}}
\right) -\frac{1}{4}\right] +\frac{1}{4}\right\}   \nonumber \\
&&+\frac{1}{2}\left( 1+\eta \frac{\sigma _{0}}{S_{0}}\right)
m_{\omega }^{2}V_{0}^{2}+\frac{1}{4!}\zeta g_{\omega
N}^{4}V_{0}^{4}+\frac{1}{2} m_{\phi }^{2}\phi
_{0}^{2}-\frac{1}{2}m_{\sigma ^{\ast }}^{2}\sigma _{0}^{\ast
^{2}},
\end{eqnarray}

\begin{eqnarray}
s &\equiv &S/V=\sum_{i}\frac{g_{i}}{6\pi ^{2}}\int dk\frac{k^{4}}{
E_{i}^{\ast }(k)}\left\{ \frac{\beta ^{2}(E_{i}^{\ast }\left(
k\right) -\nu _{i})\exp \left[ \beta \left( E_{i}^{\ast }(k)-\nu
_{i}\right) \right] }{ \left[ \exp \left[ \beta \left( E_{i}^{\ast
}(k)-\nu _{i}\right) \right] +1
\right] ^{2}}\right.   \nonumber \\
&&\left. +\frac{\beta ^{2}(E_{i}^{\ast }\left( k\right) +\nu
_{i})\exp \left[ \beta \left( E_{i}^{\ast }(k)+\nu _{i}\right)
\right] }{\left[ \exp \left[ \beta \left( E_{i}^{\ast }(k)+\nu
_{i}\right) \right] +1\right] ^{2}}. \right\}
\end{eqnarray}

In the previous calculation\cite{Yang:2003up}, we employed the
parameter set T1 given by ref.\cite{Zhang:2000ug}, and the earlier
data of strong Y-Y interaction. In refs.\cite{Song:2003xe,
Qian:2004td}, we have the new data of weak Y-Y interaction and
adjusted the parameters from $g_{\sigma^{\ast}\Lambda}^2 = 48.31$
and $g_{\sigma^{\ast}\Xi}^2 = 154.62$ for strong Y-Y interaction,
to $g_{\sigma^{\ast}\Lambda}^2 = 28.73$ and
$g_{\sigma^{\ast}\Xi}^2 = 129.06$ for weak Y-Y interaction.  The
details of the parameters are summarized in Table 1.  We will use
the parameters for weak Y-Y interaction to address the L-G phase
transition in the next section.

\begin{table}
\begin{tabular}[t]{ccccccccc}\hline
\centering
$m_s$ & $S_0$ & $\xi$ & $\eta$ & $d$ & $g_{sN}^2$ & $g_{\omega N}^2$ &  $g_{s\Lambda}^2$ & $g_{s\Xi}^2$ \\
\hline
$509$& $90.6$&$0.0402$&$-0.496$&$2.70$&$99.3$     & $154.5$          &$37.32$         &$9.99$\\
\hline\hline
$g_{\sigma^{\ast}\Lambda}^2$(Strong)&$g_{\sigma^{\ast}\Xi}^2$(Strong)&$g_{\sigma^{\ast}\Lambda}^2$(Weak)&$g_{\sigma^{\ast}\Xi}^2$(Weak)\\
\hline
$48.31$ & $154.62$ & $28.73$ & $129.06$\\
\hline
\end{tabular}\label{table1}
\caption{The parameters used in calculation}
\end{table}

\section{L-G phase transition in SHM with weak Y-Y interaction}

In this section, we employ the extended FST model to investigate
the L-G phase transition in the SHM with weak Y-Y interaction.

Using the formulae in Sec.II and the parameters of Table 1, we
calculate the pressure and study the equation of state first.  The
results for pressure vs. baryon density $\rho_B$ curves at
$T=10MeV$ are shown in Fig.1, with weak and strong Y-Y
interactions, respectively.  We can see that the equation of state
for two cases are quite different in two aspects.  Firstly, with
weak Y-Y interaction, when $f_S>0.7$, the pressure increases with
density monotonically and there are no mechanical unstable
regions.  While with strong Y-Y interaction, there are always
mechanical unstable regions for each $f_S$.  Secondly, with weak
Y-Y interaction, the minimum of the pressure-density curve rises
monotonically with $f_S$, till $f_S>0.7$, where there is no
minimum.  While with strong Y-Y interaction, the minimum drops
with $f_S$ and reaches the lowest value on the curve of $f_S=1.3$.
When $f_S>1.3$, the pressure-density curves ascend with the
increase of $f_S$.  These features are consistent with the
behavior of energy-density curves and the stability of SHM in
ref.\cite{Qian:2003ct} and \cite{Qian:2004td}.

To discuss the chemical instability of SHM with weak Y-Y
interaction, we show the chemical potential isobars for nucleons
and $\Xi$ against $f_S$ at temperature $T=10MeV$ in Fig.2 for $p =
0.05, 0.10, 0.20, 0.40, 0.50$ and $0.60 MeV fm^{-3}$ respectively.
An inflection point could be spotted on the curve with the
pressure $p^i=0.50MeVfm^{-3}$.  There are chemical unstable
regions on curves with $p < p^i$.  This behavior is similar to
that of strong Y-Y interaction but the inflection points are
different for these two cases.

The chemical, thermal and mechanical Gibbs equilibrium conditions
for two separate phases require
\begin{equation}
T^L=T^G=T
\end{equation}
\begin{equation}
\mu _q^L\left( T,\rho^L, f_s^L\right) =\mu _q^G\left( T,\rho^G,
f_s^G\right), (q=N,\Xi),
\end{equation}
\begin{equation}
p^L\left( T,\rho^L,f_s^L\right) =p^G\left( T,\rho^G,f_s^G\right),
\end{equation}
where the superscripts $L$, $G$ denote the liquid and gas phases,
respectively.  We follow the procedures used in
refs.\cite{Muller:1995ji,Qian:2000fq,Qian:2001dw,Qian:2002rp,Yang:2003up}
to solve out Eqs.(14),(15),(16).  Collecting up all the solutions
and we get the phase boundary, i.e., binodal surface.  The binodal
surface at $T=10MeV$ are shown in Fig.3, where the dotted line
refers to weak Y-Y interaction and the solid line to strong Y-Y
interaction.  We see from Fig.3 that the differences between two
binodal surfaces are very remarkable.  Though the limit pressures,
above which the L-G phase transition cannot take place, exist for
both two cases, the values are very different.  The limit pressure
for weak Y-Y interaction $p_{lim}^W=0.50MeVfm^{-3}$ is much higher
than that of the strong Y-Y interaction as
$p_{lim}^S=0.095MeVfm^{-3}$.  The enclosed region of binodal
surface with weak Y-Y interaction is much larger than that of the
strong Y-Y interaction.  A higher limit pressure and a larger
region of phase separation imply that the system is likely to be
less stable and the L-G phase transition can take place more
easily. This result is of course very reasonable if we notice the
conclusion given by refs\cite{Song:2003xe,Qian:2004td} that the
SHM with weak Y-Y interaction is less stable.

To determine the order of the L-G phase transition, we examine the
entropy density and the specific heat capacity of SHM with weak
Y-Y interaction.  By Ehrenfest's definition, the first order phase
transition is characterized by the discontinuities of the first
order derivatives of the chemical potential, such as the
discontinuities of entropy and volume, while the second order
phase transition unfolds the discontinuous behavior for the second
order derivatives of the chemical potential, such as the specific
heat capacity. The entropy per baryon is
\begin{equation}
s(T,p,f_{s})=\frac{S(T,p,f_{s})}{\rho_{B}}.
\end{equation}
where $S(T,p,f_{s})$ can be calculated from Eq.(13).

The specific heat capacity is
\begin{eqnarray}
C_{p} &=& T(\frac{\partial S}{\partial T})_{p,f_{s}}\nonumber \\
&&= T \sum_{i}\frac{g_{i}}{6\pi ^{2}}\int dk\frac{k^{4}}{
E_{i}^{\ast }(k)} \left[\frac{d^{2}n_{i}}{dT^{2}} +
\frac{d^{2}\overline{n}_{i}}{dT^{2}} \right],
\end{eqnarray}
where
\begin{eqnarray}
\frac{d^{2}n_{i}}{dT^{2}}= \frac{ e^{  \beta \left( E_{i}^{\ast
}(k)-\nu _{i}\right) }\left[( E_{i}^{\ast }(k)-\nu _{i})^{2}(e^{
\beta \left( E_{i}^{\ast }(k)-\nu _{i}\right) } - 1)-2T(
E_{i}^{\ast }(k)-\nu _{i}) (e^{\beta \left( E_{i}^{\ast }(k)-\nu
_{i}\right)}+1)\right]
 }{T^{4}(e^{\beta \left( E_{i}^{\ast
}(k)-\nu _{i}\right)}+1)^{3}},
\end{eqnarray}
\begin{eqnarray}
\frac{d^{2}\overline{n}_{i}}{dT^{2}}= \frac{ e^{  \beta \left(
E_{i}^{\ast }(k)+\nu _{i}\right) }\left[( E_{i}^{\ast }(k)+\nu
_{i})^{2}(e^{ \beta \left( E_{i}^{\ast }(k)+\nu _{i}\right) } -
1)-2T( E_{i}^{\ast }(k)+\nu _{i}) (e^{\beta \left( E_{i}^{\ast
}(k)+\nu _{i}\right)}+1)\right]
 }{T^{4}(e^{\beta \left( E_{i}^{\ast
}(k)+\nu _{i}\right)}+1)^{3}}.
\end{eqnarray}

The entropy vs. temperature curves for $f_S=0.0,0.1$ and $0.4$ at
$p=0.11MeVfm^{-3}$ are shown in Fig.4.  We see from Fig.4 that the
entropy evolves continuously through the phase transition when
$f_S>0.0$. But when $f_S=0.0$, our model reduces to the
one-component symmetric nuclear matter.  The entropy becomes
discontinuous.  The L-G phase transition becomes the first order.
 This result is consistent with theoretical and experimental results about L-G
phase transition in symmetric nuclear matter.

The specific heat capacity vs. temperature curves for $f_S=0.1$
and $0.4$ at $p=0.11MeVfm^{-3}$ are shown in Fig.5.  The specific
heat capacity curves are discontinuous for $f_S=0.1$ and $0.4$.
 The continuity of entropy together with the discontinuity of
specific heat capacity demonstrates that the L-G phase transition
in SHM with weak Y-Y interaction is still of second order, as
expected.

One of the important signals for L-G phase transition is the
investigation of Caloric
curves\cite{Kolomietz:2001gd,Ma:2003dc,Ma:2004ey,Sil:2004bu}. The
Caloric curve reflects the relationship between excited energy of
hadronic matter and the temperature.  The energy of SHM is
\begin{eqnarray}
\varepsilon &=&\sum_{i}\frac{g_{i}}{\pi ^{2}}\int dk
k^{2}E_{i}^{\ast }(k) [n_{i}(k)+\overline{n}_{i}(k)]+H_{q}\left\{
\left( 1-\frac{\sigma _{0}}{S_{0} }\right) ^{\frac{4}{d}}\left[
\frac{1}{d}ln\left( 1-\frac{\sigma _{0}}{S_{0}}
\right) -\frac{1}{4}\right] +\frac{1}{4}\right\}   \nonumber \\
&&-\frac{1}{2}\left( 1+\eta \frac{\sigma _{0}}{S_{0}}\right)
m_{\omega }^{2}V_{0}^{2}-\frac{1}{4!}\zeta g_{\omega
N}^{4}V_{0}^{4}-\frac{1}{2} m_{\phi }^{2}\phi
_{0}^{2}+\frac{1}{2}m_{\sigma ^{\ast }}^{2}\sigma _{0}^{\ast ^{2}}
\nonumber \\
&&+g_{\omega
N}V_{0}\rho_{BN}+(g_{\omega\Lambda}V_{0}+g_{\phi\Lambda}\phi_{0})\rho_{B\Lambda}+(g_{\omega\Xi}V_{0}+g_{\phi\Xi}\phi_{0})\rho_{B\Xi}
\end{eqnarray}
and the excited energy $E^{\ast}/\rho_B$ is\cite{Kolomietz:2001gd}
\begin{eqnarray}
E^{\ast}/\rho_B = \frac{\lambda\varepsilon^{L}(\rho^L, fs^L,
T)+(1-\lambda)\varepsilon^{G}(\rho^G, fs^G,
T)}{\lambda\rho^{L}+(1-\lambda)\rho^{G}}-\left(\frac{\varepsilon^{L}(\rho^L,
fs^L, T)}{\rho^{L}}\right)_{T=0}
\end{eqnarray}

We present the Caloric curves at pressure $p=0.11MeVfm^{-3}$ for
$f_S=0.0,0.1$ and $0.4$ in Fig.6.  There are slopes during the L-G
phase transition for $f_S=0.1$ and $0.4$, while for $f_S=0.0$
there is a platform.  This platform-shaped Caloric curve has been
got in experiments of the L-G phase transition in normal symmetric
nuclear matter\cite{Ma:2003dc,Ma:2004ey}.  For $f_S>0.0$, the
slopes mean that temperature does not remain constant during phase
transition.  This is the very feature of second order phase
transition as was first pointed out by M\"{u}ller and
Serot\cite{Muller:1995ji}.

To exhibit the behavior of Caloric curves of SHM with strong and
weak Y-Y interactions clearly, we plot the temperature $T$ vs.
$E^{\ast}/\rho_B$ curves at $p=0.11MeVfm^{-3}$, $f_S=0.4$ for
strong and weak Y-Y interactions in Fig.7.  We find the
temperature region of phase transition for weak Y-Y interaction is
lower than that of strong Y-Y interaction and the slope for weak
Y-Y interaction is more acclivitous.  Perhaps these distinctions
can provide us a possibility to check the Y-Y interaction be
strong or weak, and then the $\Lambda - \Lambda$ energy.

\section{Summary}
We have used the new parameters about the weak Y-Y interaction
derived from recent experiment data to reexamine the L-G phase
transition in SHM.  We have employed the extended FST model with
nucleons and $\Lambda$, $\Xi$ hyperons and simplified the SHM into
a two-component system to discuss its L-G phase transition.  We
come to the following conclusions:
\begin{enumerate}
\item No matter the Y-Y interaction is strong or weak, the L-G
phase transition can take place in SHM.  The L-G phase transition
is of second order for $f_S\neq0$, the entropy is continuous and
the heate capacity is discontinuous at the phase transition point.
But for $f_S=0$ the L-G phase transition becomes the first order.
The entropy is discontinuous.

\item Due to the difference between the stability of SHM with
strong and weak Y-Y interactions, the physical features of L-G
phase transition, including the limit pressure, the regions
enclosed by the binodal surface, the Caloric curves, the slope of
phase transition curves are very different for weak and strong Y-Y
interactions.

\item After reexaming the features of SHM, we come to a conclusion
that the L-G phase transition can occur in SHM with weak Y-Y
interaction more easily than that with strong Y-Y interaction.

\end{enumerate}

\section{Acknowledgements}
This work is supported in part by National Natural Science
Foundation of China under Nos. 10235030, 10247001, 10375013,
10347107 10047005, 10075071, the National Basic Research Programme
2003CB716300, the Foundation of Education Ministry of China under
contract 2003246005 and CAS Knowledge Innovation Project No.
KJCX2-N11.

\begin{figure}[tbp]
\includegraphics[totalheight=10cm, width=8cm]{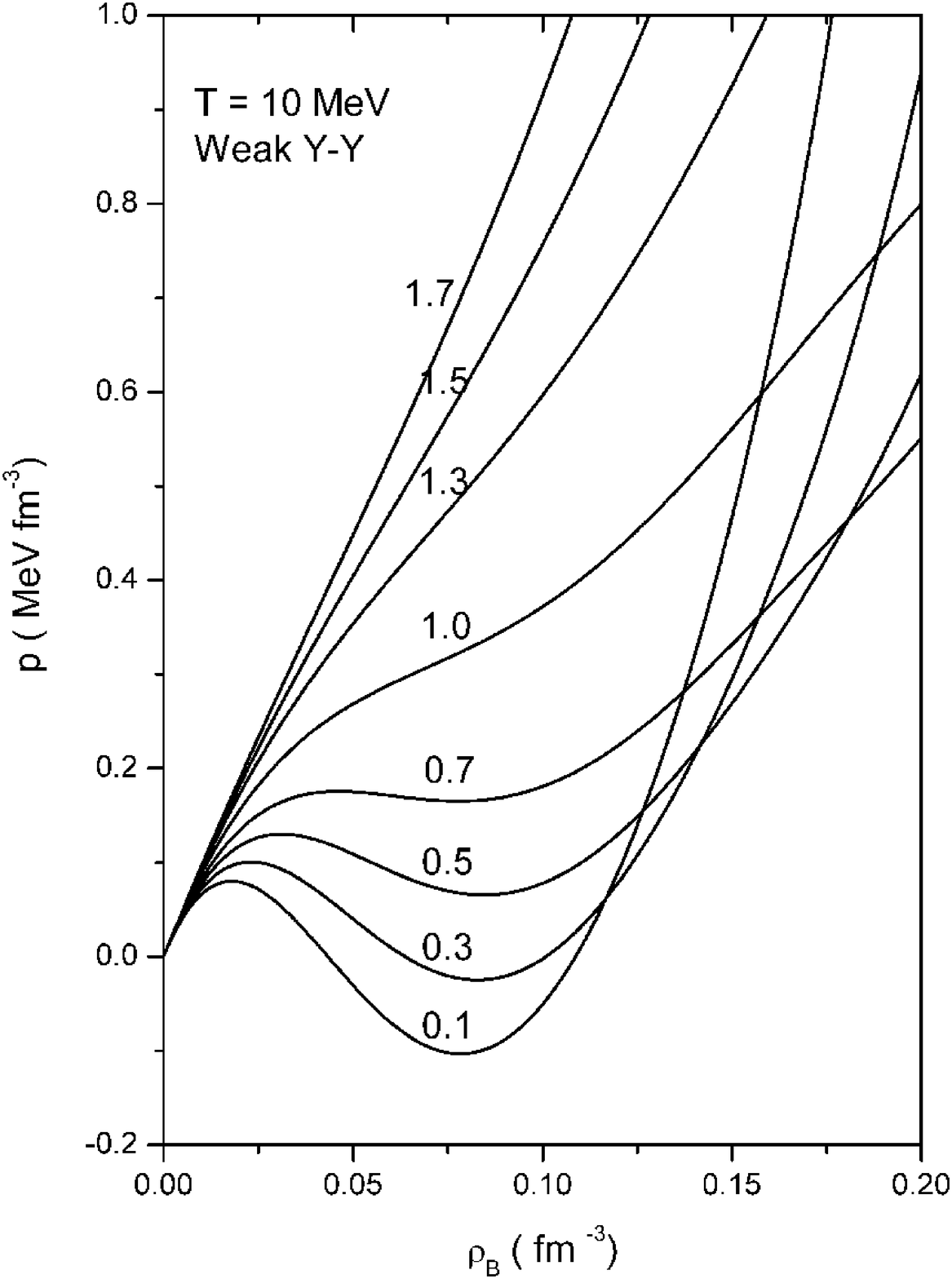}
\includegraphics[totalheight=10cm, width=8cm]{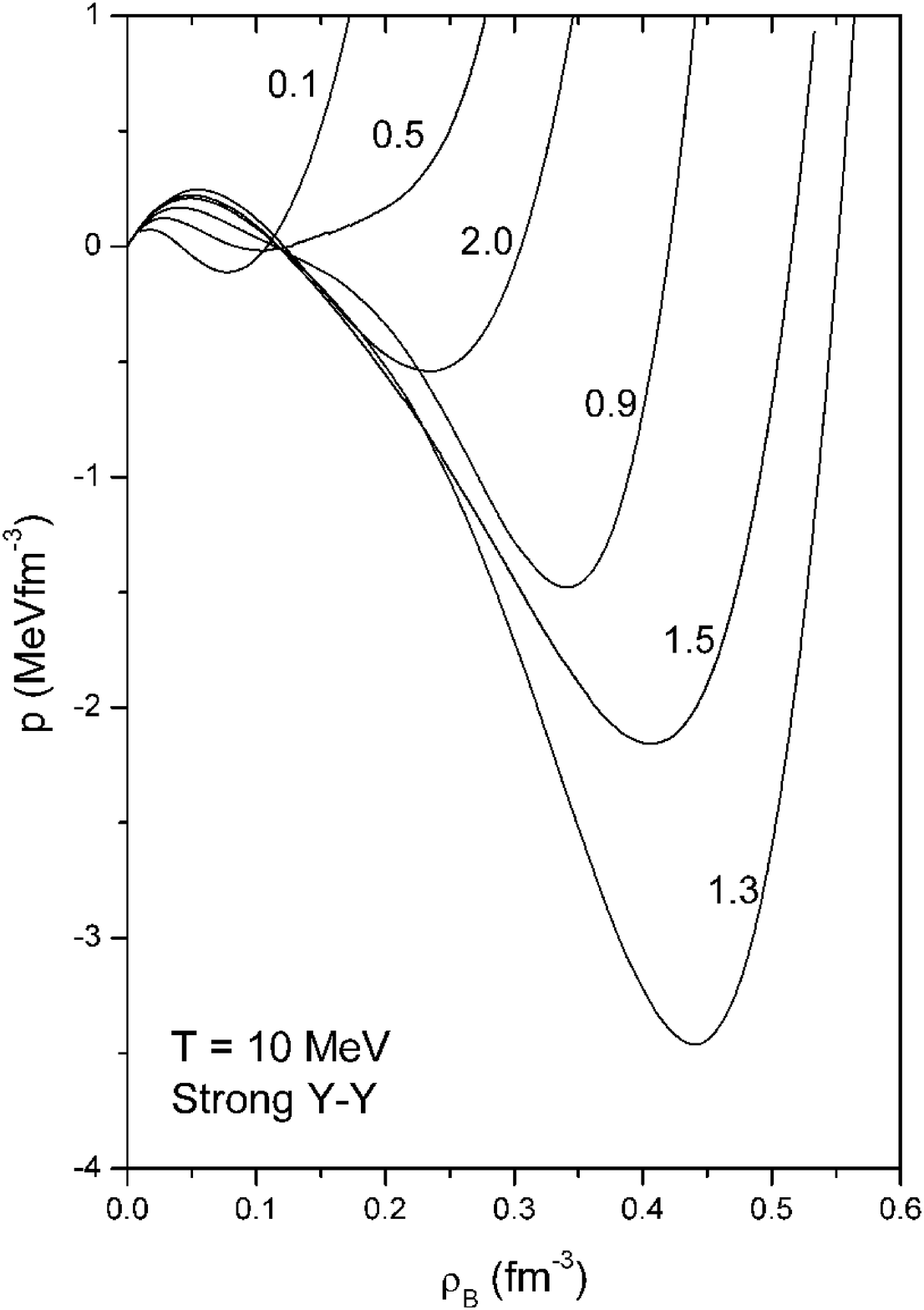}

\caption{Pressure as a function of baryon density at temperature
$T = 10 MeV$ for various strangeness fractions $f_{S}$ with weak
and strong Y-Y interactions.  Note that the scales of the two
figures are not the same.} \label{fig1}
\end{figure}

\begin{figure}[tbp]
\includegraphics[totalheight=16cm, width=16cm]{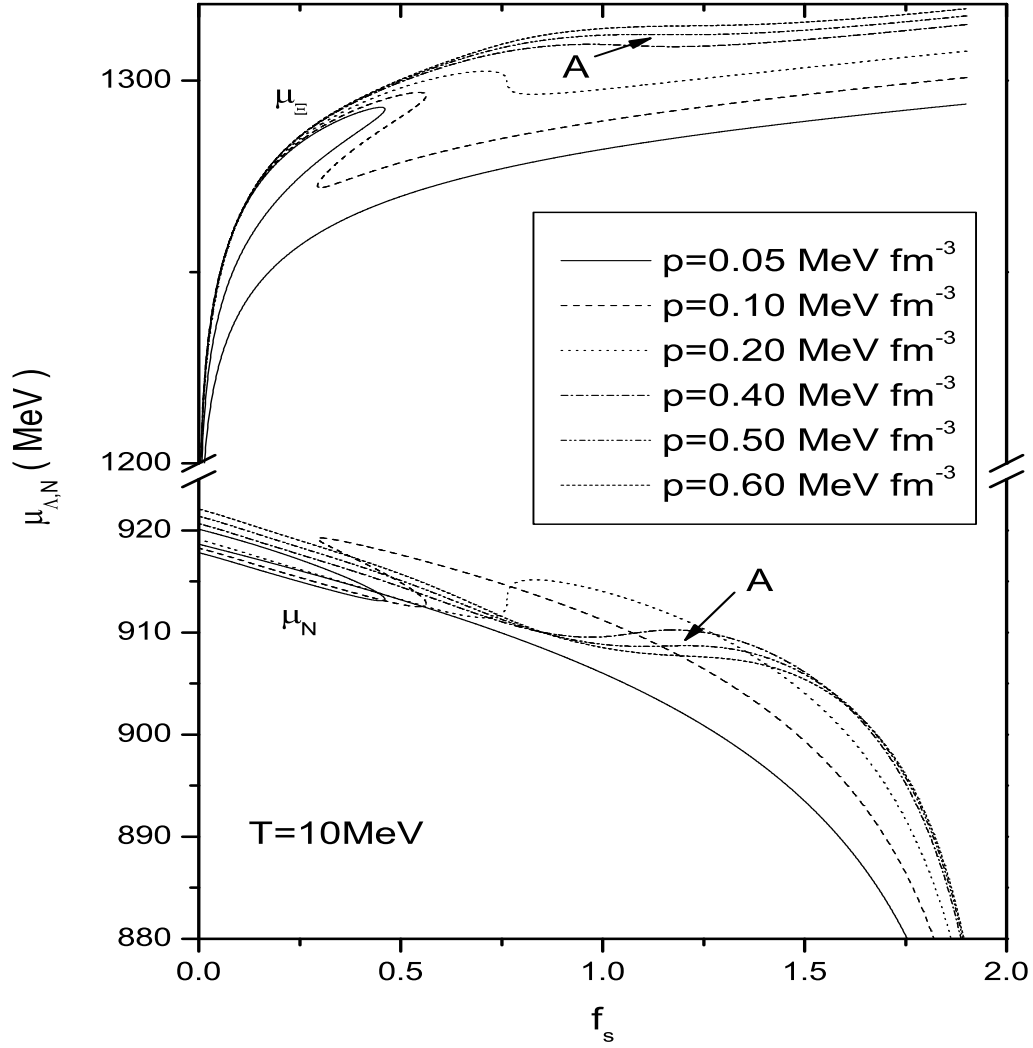}
\caption{Chemical potential isobars as functions of $f_{S}$ at $T
= 10 MeV$. The curves have pressures $p = 0.05, 0.10, 0.20, 0.40,
0.50$ and $0.60 MeV fm^{-3}$ respectively.  The inflection
point(A) is at $p^i=0.50MeVfm^{-3}$.} \label{fig2}
\end{figure}

\begin{figure}[tbp]
\includegraphics[totalheight=16cm, width=16cm]{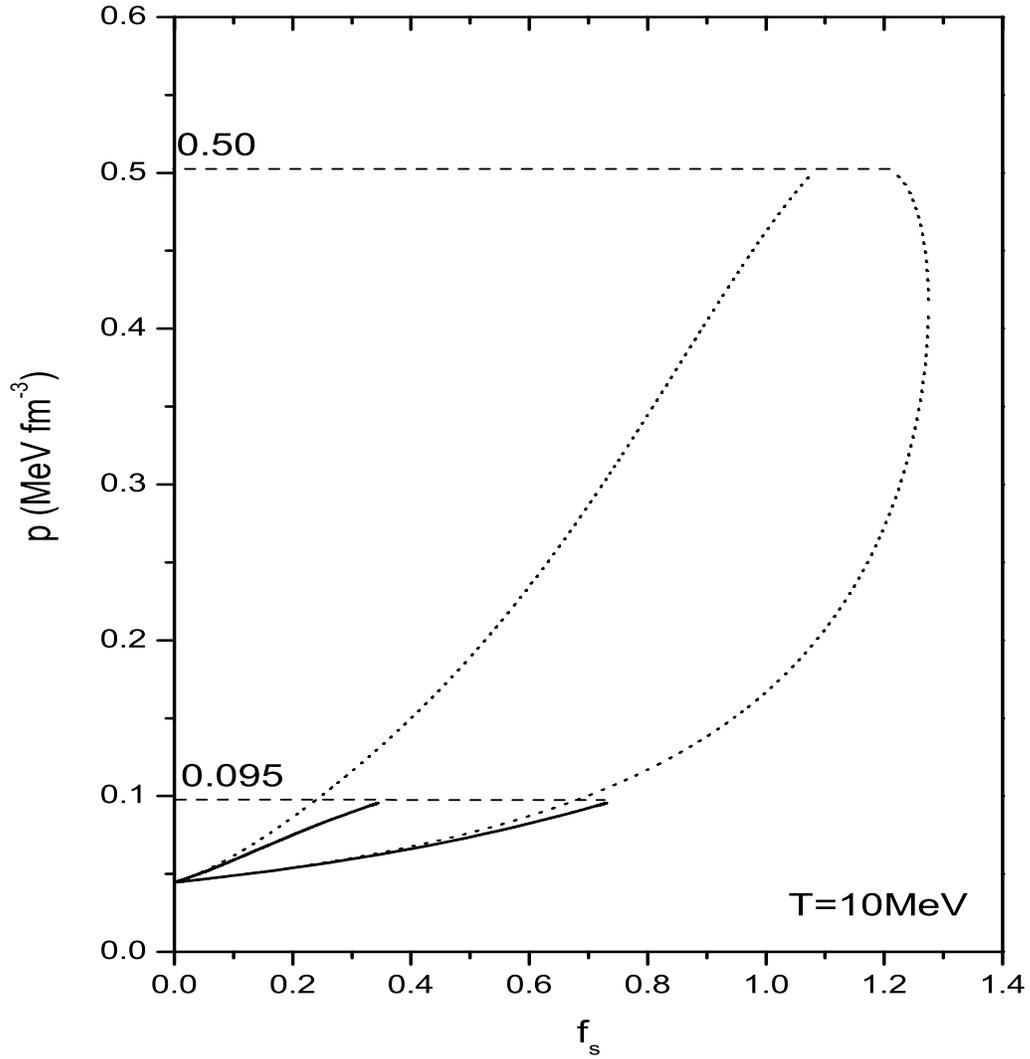}
\caption{Binodal surface at $T = 10 MeV$. The dotted line is the
biondal surface for weak Y-Y interaction, while the solid line is
that for strong Y-Y interaction. The binodal surfaces are cut off
at limit pressure $p_{lim}=0.50MeVfm^{-3}$ and
$p_{lim}=0.095MeVfm^{-3}$ respectively.}
 \label{fig3}
\end{figure}

\begin{figure}[tbp]
\includegraphics[totalheight=16cm, width=16cm]{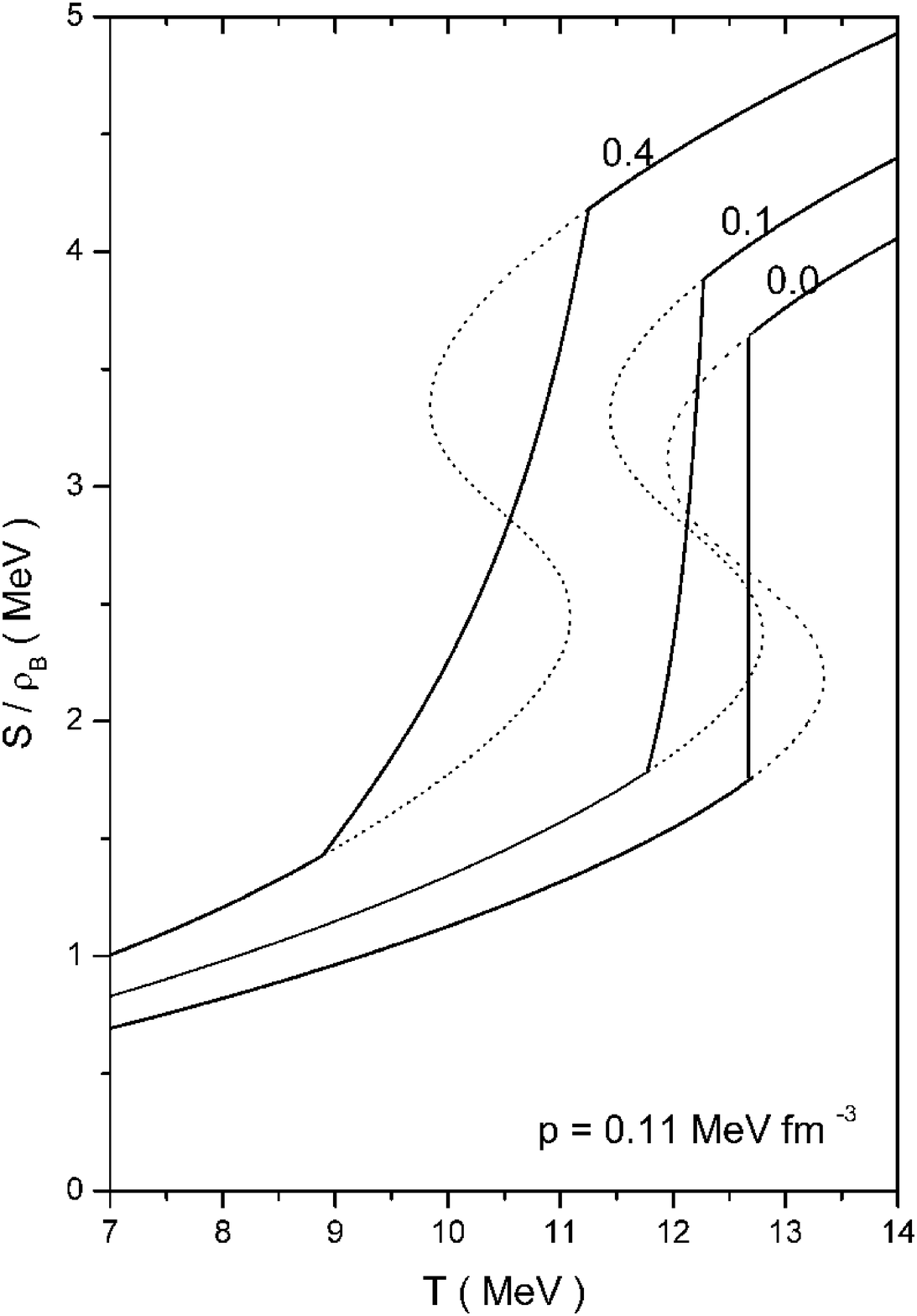}
\caption{Entropy as a function of temperature at constant pressure
$p=0.11MeVfm^{-3}$ for strangeness fraction $f_S=0.0,0.1$ and
$0.4$. For $f_S=0.1$ and $0.4$, the entropy evolves continuously
through the phase transition, while for $f_S=0.0$ the entropy is
discontinuous.} \label{fig4}
\end{figure}

\begin{figure}[tbp]
\includegraphics[totalheight=16cm, width=16cm]{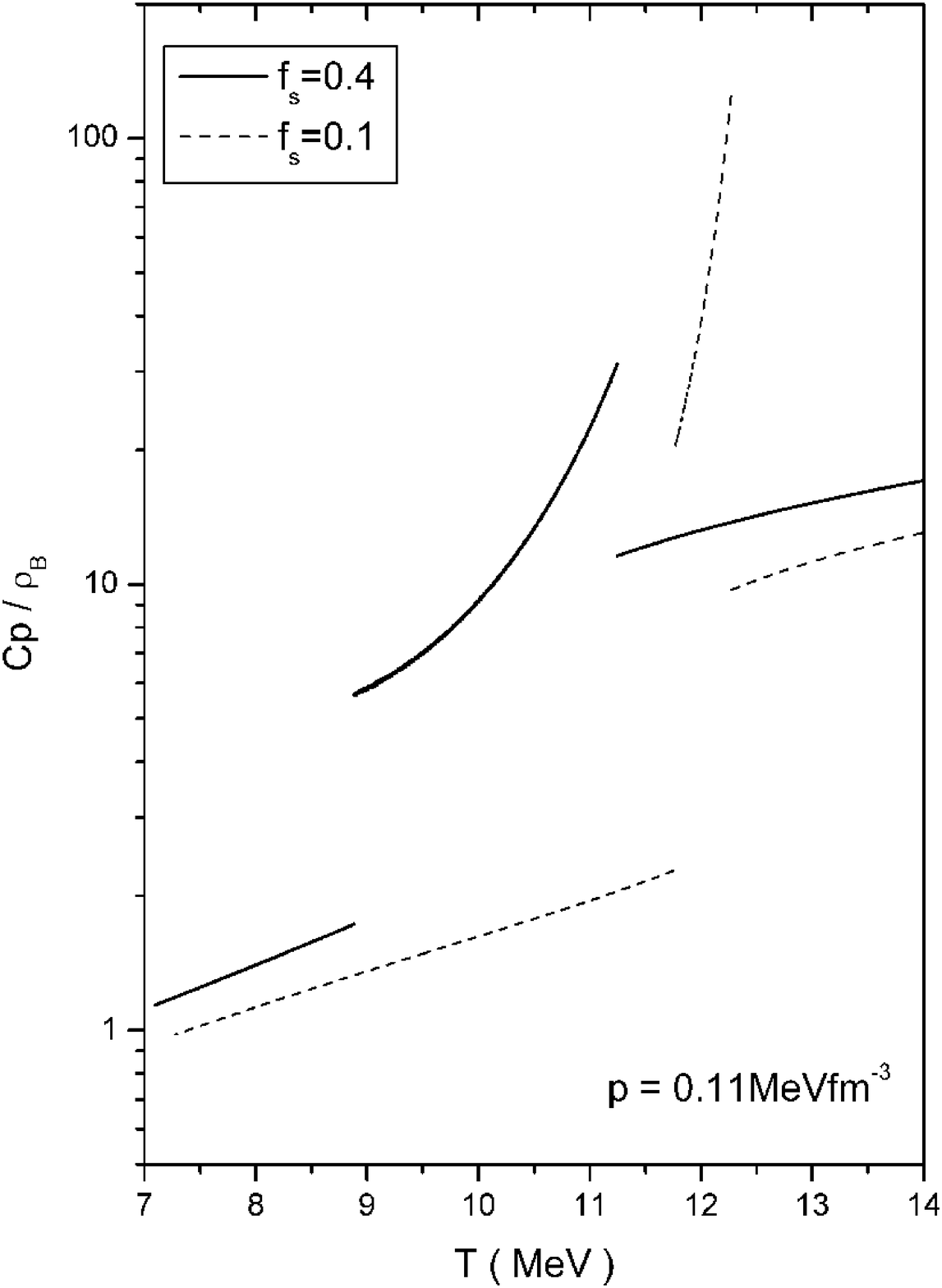}
\caption{Specific heat capacity as a function of temperature at
constant pressure $p=0.11MeVfm^{-3}$ for strangeness fraction
$f_S=0.1$ and $0.4$. For $f_S=0.1$ and $0.4$, the entropy is
discontinuous through the phase transition.} \label{fig5}
\end{figure}

\begin{figure}[tbp]
\includegraphics[totalheight=16cm, width=16cm]{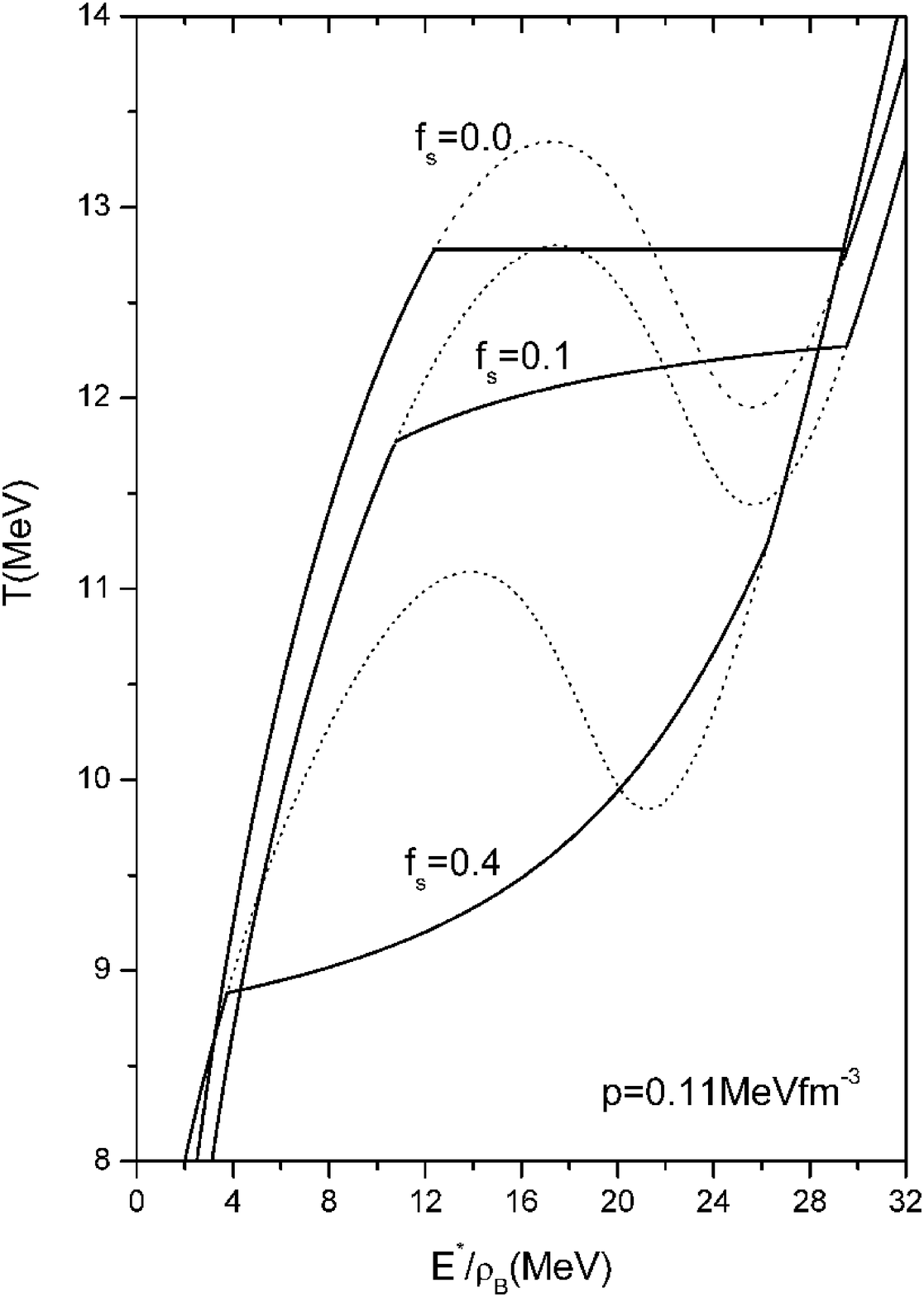}
\caption{Temperature as a function of excited energy at constant
pressure $p=0.11MeVfm^{-3}$ for strangeness fraction $f_S=0.0,0.1$
and $0.4$. For $f_S=0.1$ and $0.4$, there are slopes during the
phase transition, while for $f_S=0.0$ there is a platform.}
\label{fig6}
\end{figure}

\begin{figure}[tbp]
\includegraphics[totalheight=16cm, width=16cm]{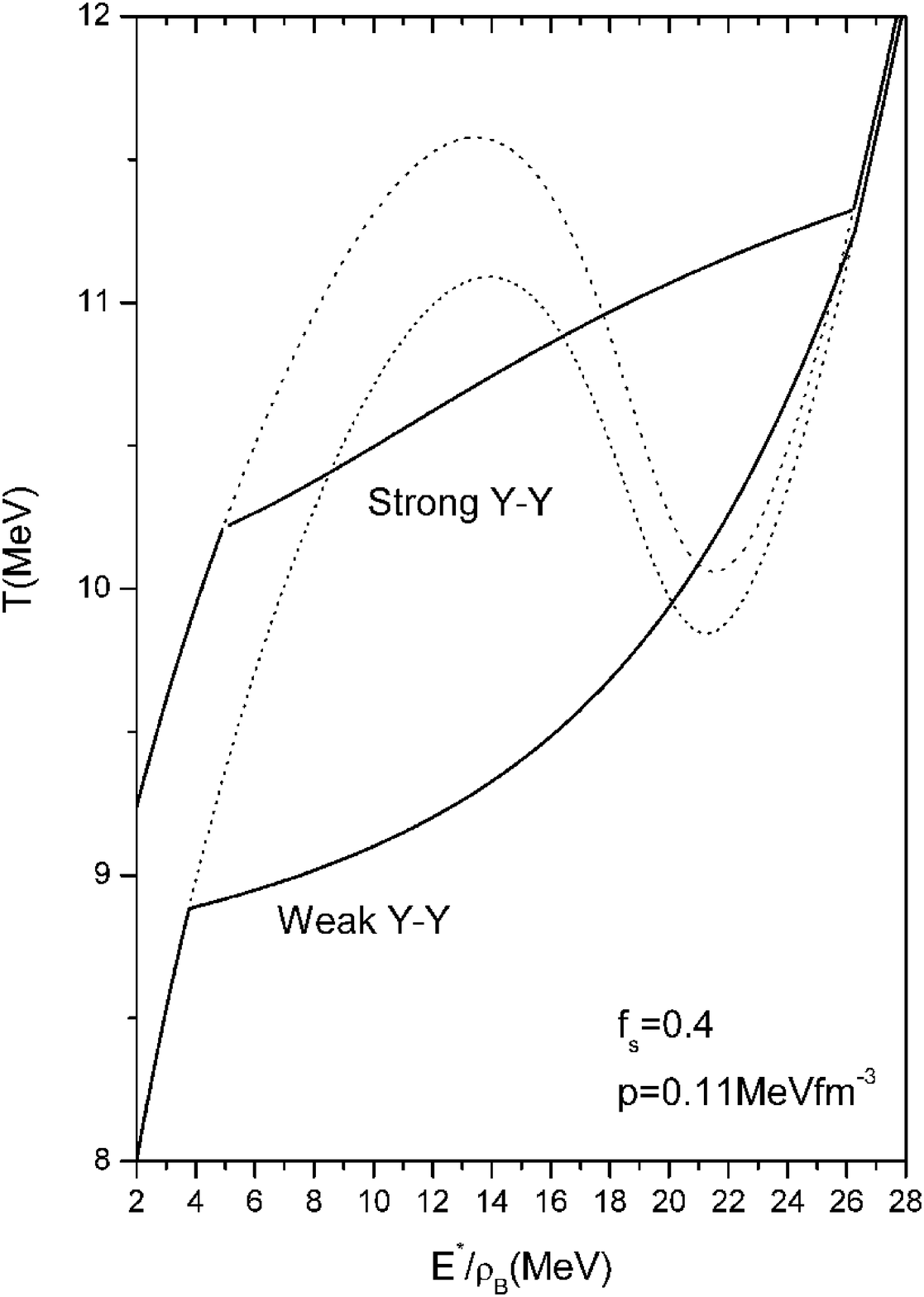}
\caption{Temperature as a function of excited energy at constant
pressure $p=0.11MeVfm^{-3}$ for strangeness fraction $f_S=0.4$.}
\label{fig7}
\end{figure}

\end{document}